\documentclass{desyproc}
\begin{document}


\title{J/$\psi$ production as a function of charged particle multiplicity in pp collisions at $\sqrt{s}=7$ TeV with the ALICE experiment.}



%
%
%
%
%
%
%
%
%

\author{{\slshape Sarah Porteboeuf-Houssais}  for the ALICE Collaboration\\[1ex]
Laboratoire de Physique Corpusculaire (LPC), Clermont Universit\'e, Universit\'e Blaise Pascal, CNRS-IN2P3, Clermont-Ferrand, France}


\contribID{ZZ}
\confID{UU}
\desyproc{DESY-PROC-2012-YY}
\acronym{MPI@LHC 2011}
\doi
\maketitle


\begin{abstract}
We report on the first measurement of J/$\psi$ production as a function of charged particle pseudo-rapidity density d$N_\textrm{ch}/d\eta$ in pp collisions at $\sqrt{s}$ = 7 TeV with the ALICE experiment at the LHC. J/$\psi$ mesons are detected down to $p_{\rm{T}}$ = 0 via their decays into $e^{+}e^{-}$ pairs at mid-rapidity ($|y|$ $<$ $0.9$) and into $\mu^{+} \mu^{-}$  pairs at forward rapidity ($2.5<y<4$).  d$N_\textrm{ch}/d\eta$ is measured within $|\eta| < 1$. We compare results in the two different J/$\psi$ rapidity ranges. Preliminary PYTHIA simulations are also presented. 
\end{abstract}



\section{Introduction}

The production mechanism of heavy quarkonium states (eg. $J/\psi$) is very complex and is not fully understood. Various models such as the Color Singlet, nonrelativistic QCD approach (NRQCD) and the Color Evaporation Model aim to explain how a heavy resonance state can be produced in a hard process \cite{quarkonia_review, Lansberg}. This field is very active  in theory development. 
In particular, describing the J/$\psi$ production cross-section and polarization is a challenge for most of the models \cite{Butenschoen_pt, Butenschoen_pol, Lansberg_ptpol}, including also the last LHC data \cite{alice_Jpsi, atlas_Jpsi, cms_Jpsi, lhcb_Jpsi}. 
Furthermore, J/$\psi$ production could be accompanied by a hadronic activity (hadrons produced in a cone around J/$\psi$) and it was pointed out that new observables are needed to constrain models \cite{ConesadelValle}. In addition, it was proposed in \cite{Frankfurt,Strikman1,Strikman2} that intial state effects could modify J/$\psi$ production due to gluon density fluctuations and a special transverse structure of the nucleon.

\smallskip
To look at exclusive final states and not only inclusive ones, it is needed to have a full description of  hard processes in a complete event. A description of the interplay between the hard and the soft components of the event as well as of color flow and energy conservation is mandatory. In high energy proton-proton collisions, the total event multiplicity can have a substantial contribution from Multi-Parton Interactions (MPI). With MPI, several parton-parton interactions can occur in a single pp collision. MPI are commonly used to describe the soft underlying event  but can also contribute on the hard and semi-hard scale, this contribution becoming more and more relevant with increasing energy \cite{MPI_gribovregge, Sjostrand_MPI, MPI_perugia}.  
The NA27 experiment performed a study that related open charm production and underlying event properties for pp collisions at $\sqrt{s}=27$ GeV. It was found that charged particle multiplicity distribution in events with and without charm production differs by $~20\%$ \cite{NA27}, indicating already a different behavior in multiplicity distribution.

\smallskip
In addition to MPI, other non-trivial effects could have an impact on multiplicity dependence and hard interactions. The measured charged particle multiplicty in pp collisions at LHC energies reaches values higher than in peripheral Cu-Cu at RHIC at $\sqrt{s_{NN}}=200$ GeV \cite{Phobos_CUCU}. With the high energy density reached in pp collisions at LHC energies, some models predict the occurence of collective behaviour at LHC energies \cite{Kisiel_collectivity_pp, Werner_collectivity_pp}, and one could consider the possibility of a modification of J/$\psi$ yield in high multiplicity pp events due to collective phenomena \cite{Vogel_quarkonia_pp}.

\smallskip
In this article, we report the measurement of relative J/$\psi$ production  (d$N_{\textrm{J}/\psi}$ / d$y$) /  $<$d$N_{\textrm{J}/\psi}/$d$y>$ at mid ($|y|<0.9$) and forward ($2.5<y<4$) rapidities as a function of the relative charged particle multiplicity (d$N_{\textrm{ch}}$/d$\eta$)/$<$d$N_{\textrm{ch}}$/d$\eta>$ for pp collisions at $\sqrt{s}=7$ TeV at LHC measured by the ALICE experiment \cite{jpsimult_paper}. We will then discuss a preliminary comparison with PYTHIA 6.4 simulations.

\section{ J/$\psi$ production as a function of the relative charged particle multiplicity}

A detailed description of the ALICE setup can be found in \cite{Alice_setup}. The J/$\psi$ is detected in the di-electron channel with the central barrel ($|\eta|<0.9$) and in the di-muon channel with the muon spectrometer ($-4<\eta<-2.5$ \footnote{In the official ALICE reference frame the muon spectrometer is located at negative $z$ positions and thus negative (pseudo-)rapidities. Since pp collisions are symmetric relative to $y=0$, we have dropped the minus sign when rapidities are quoted.}). The description of the part of the ALICE setup used in this analysis can be found in \cite{jpsimult_paper}.

\begin{figure}[!t]
\begin{center}
\includegraphics[scale=0.47]{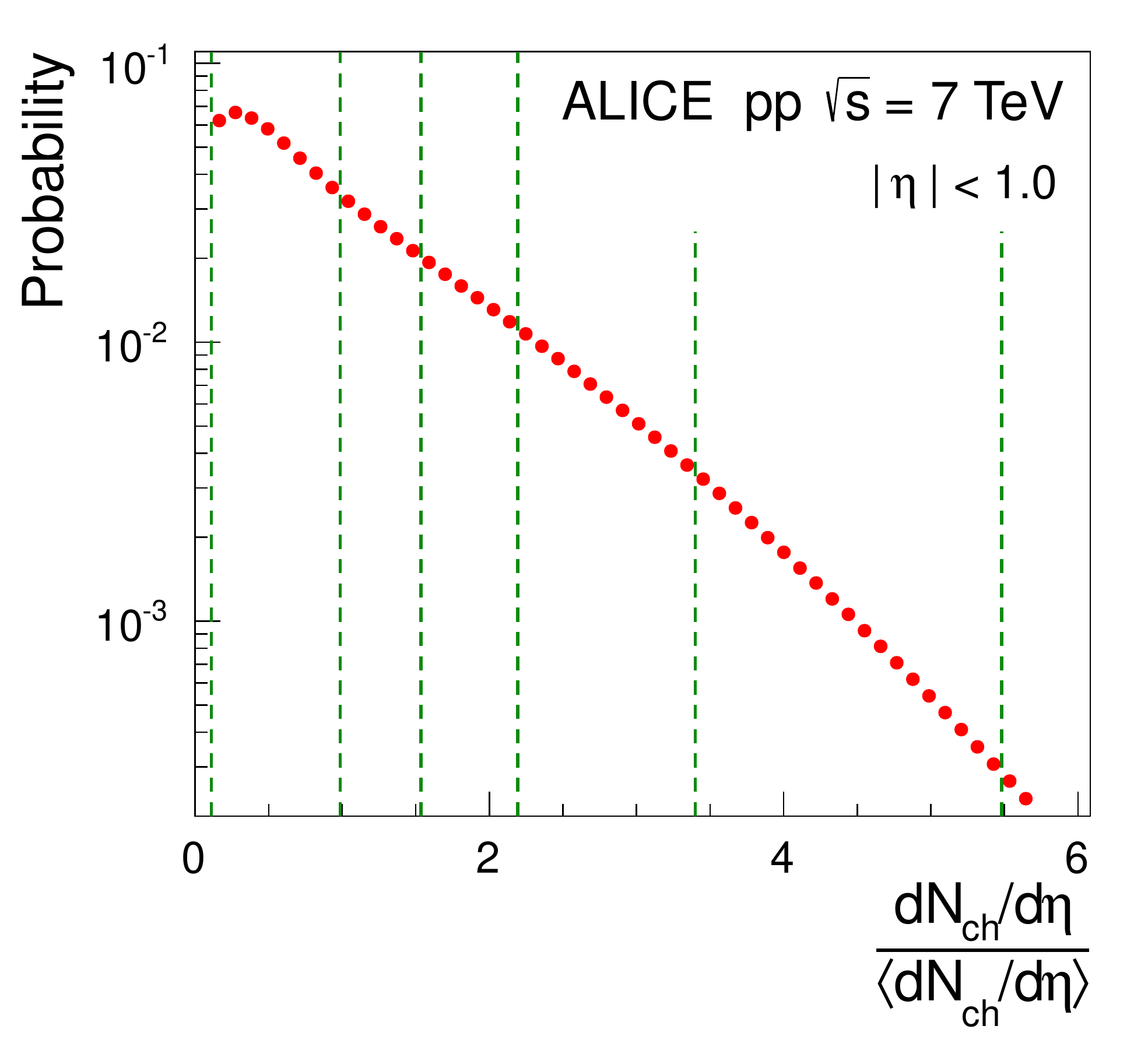}
\caption{\label{mult} Distribution of the relative charged particle density (d$N_{\textrm{ch}}$/d$\eta$)/$<$d$N_{\textrm{ch}}$/d$\eta>$ at mid-rapidity ($|\eta|<1.0$) after correction for SPD inefficiencies. Vertical lines indicate the boundaries of multiplicity bins considered in this article.}
\end{center}
\end{figure}

The results shown here were obtained by analyzing pp collisions at $\sqrt{s}=7$ TeV collected in 2010. A sample of $3.0 \times10^8$ minimum bias (MB) events and $6.75\times10^6$ $\mu$-MB triggered events were used for J/$\psi$ measurement in the di-electron and di-muon channels. This corresponds to an integrated luminosity of 4.5 nb$^{-1}$ and 7.7 nb$^{-1}$ respectively for the di-electron and di-muon channels. The MB pp trigger is defined with a signal in one of the two VZERO detectors, plus one readout chip signal in the Silicon Pixel Detector (SPD), in coïncidence with proton bunches from both sides of the interaction region. The $\mu$-MB trigger requires, in addition to the MB trigger, the detection of at least one muon with $p_{\textrm{T}}^{\textrm{trig}}>0.5$ GeV/$c$ in the acceptance of the muon arm. More details on the data sets, triggers, running conditions and relative normalization can be found in  \cite{jpsimult_paper}. Events with an interaction vertex not within $|z_{vtx}|<10$ cm are rejected. Pile-up events are identified by the presence of a second interaction vertex reconstructed in addition to the main vertex. They are rejected if the distance along the beam axis between the two vertices is larger than $0.8$ cm, and if the second vertex has at least three associated tracklets. A tracklet is defined as any combinations of two hits in the SPD layers, one hit in the inner layer and one in the outer.

\smallskip
The charged particle density d$N_{\textrm{ch}}$/d$\eta$ is estimated using the number of tracklets $N_{\textrm{trk}}$ reconstructed from hits in the SPD. Using simulated events, it was verified that $N_{\textrm{trk}}$ is proportonal to d$N_{\textrm{ch}}/$d$\eta$. Fig. \ref{mult} shows the distribution of the relative charged particle density (d$N_{\textrm{ch}}$/d$\eta$)/$<$d$N_{\textrm{ch}}$/d$\eta>$ after correction for SPD inefficiencies. 
$<$d$N_{\textrm{ch}}$/d$\eta>$ was measured for inelastic pp collisions with at least one charged particle in $|\eta|<1$ and is equal to $6.01 \pm 0.01$(stat.)$^{+0.20}_{-0.12}$(syst.) \cite{pp_mesure}. Vertical dashed lines show the limit of the $5$ bins in multiplicity used in this analysis.

    \begin{figure}[!t]
   \begin{minipage}[b]{0.45\linewidth}
\includegraphics[width=1. \columnwidth]{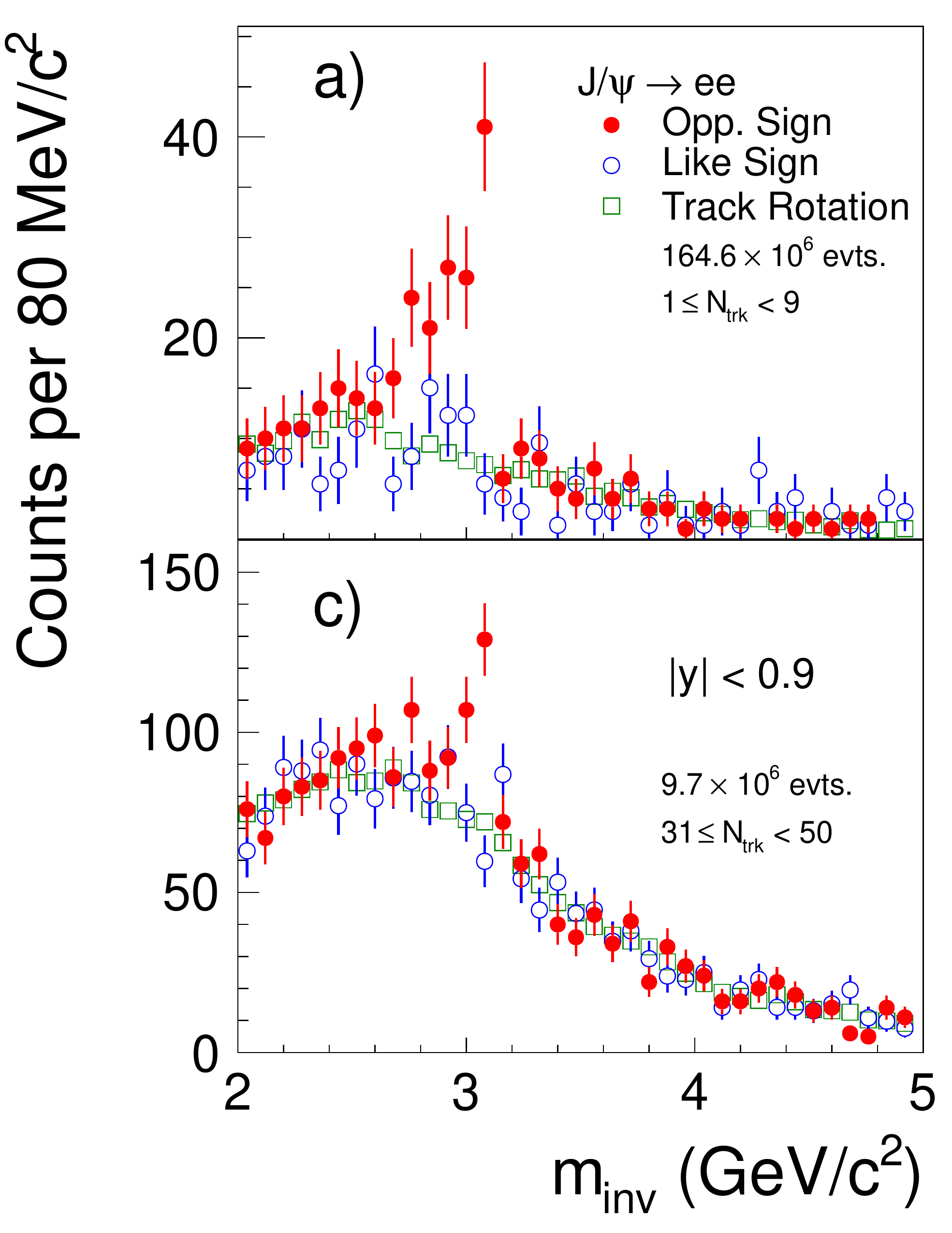}
   \end{minipage}\hfill
   \begin{minipage}[b]{0.45\linewidth}  
\includegraphics[width=1.\columnwidth]{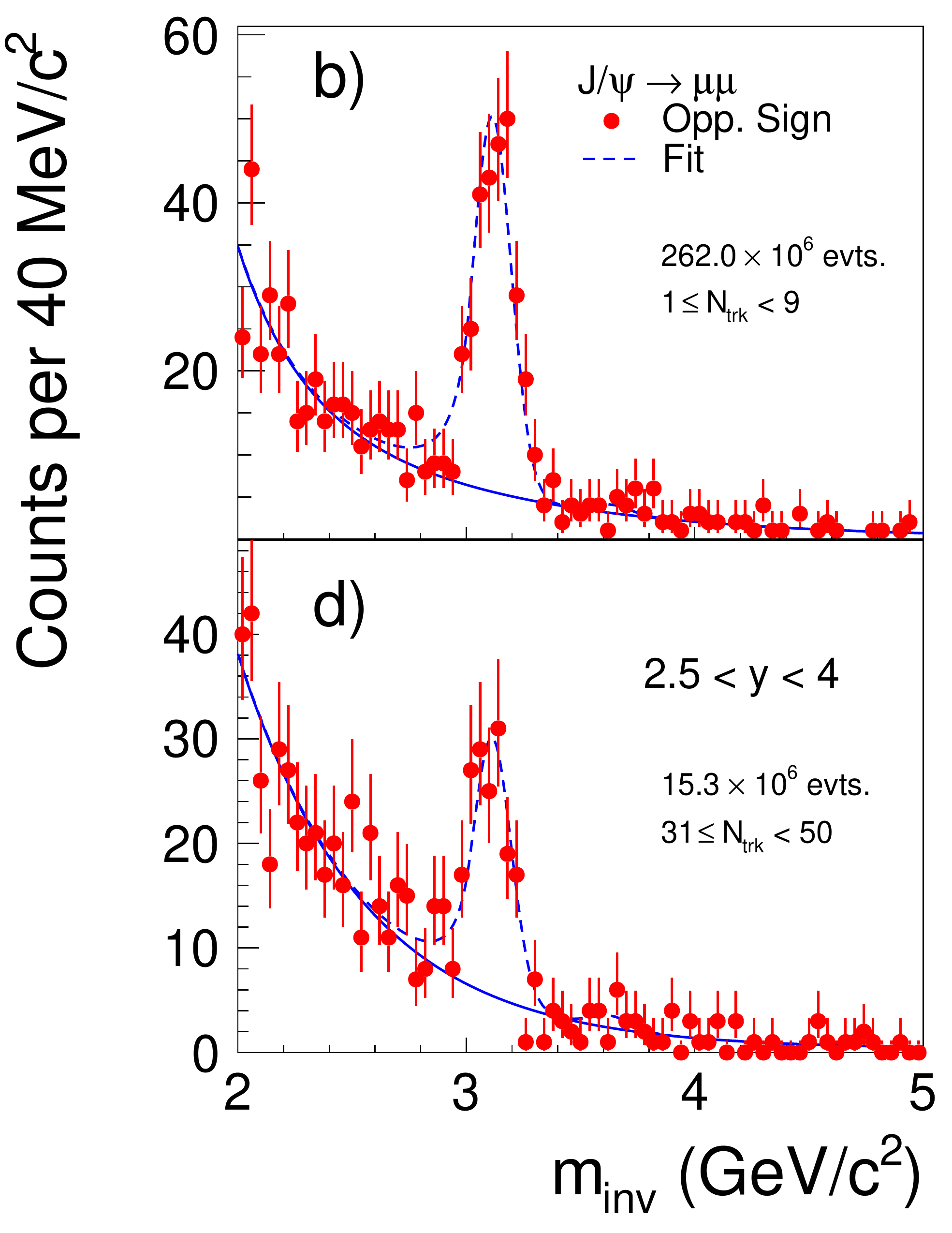}
   \end{minipage}
   \caption{Opposite sign invariant mass spectra of the selected $e^+e^-$ pairs [(a),(c)] and $\mu^+\mu^-$ pairs [(b),(d)] (full symbols) for the lowest [(a),(b)] and highest [(c),(d)] multiplicity bins. The number of events quoted in the figures refers to the corresponding minimum bias events.}
   \label{Minv}
\end{figure}

For J/$\psi$ measurement in the di-electron channel, tracks are selected by requiring transverse momentum $p_{\rm{T}}>1$ GeV/c and a pseudo-rapidity cut of $|\eta|<0.9$. Particle identification is performed by measuring the specific energy deposit $dE/dx$ in the Time Projection Chamber (TPC) of the central barrel. The invariant mass distributions of $e^+e^-$ pairs are measured in intervals of the charged particle mulitiplicity as measured via the SPD tracklets. Examples of such mass distributions are shown, for the lowest and highest multiplicity intervals in the two left panels of Fig. \ref{Minv}. The track rotation method (green squares in Fig. \ref{Minv}, left panel) is used to describe the combinatorial background in each multiplicity intervals as well as the like sign distributions (open blue circles in Fig. \ref{Minv}, left panel). 

\begin{figure}[!t]
\begin{center}
\includegraphics[scale=0.50]{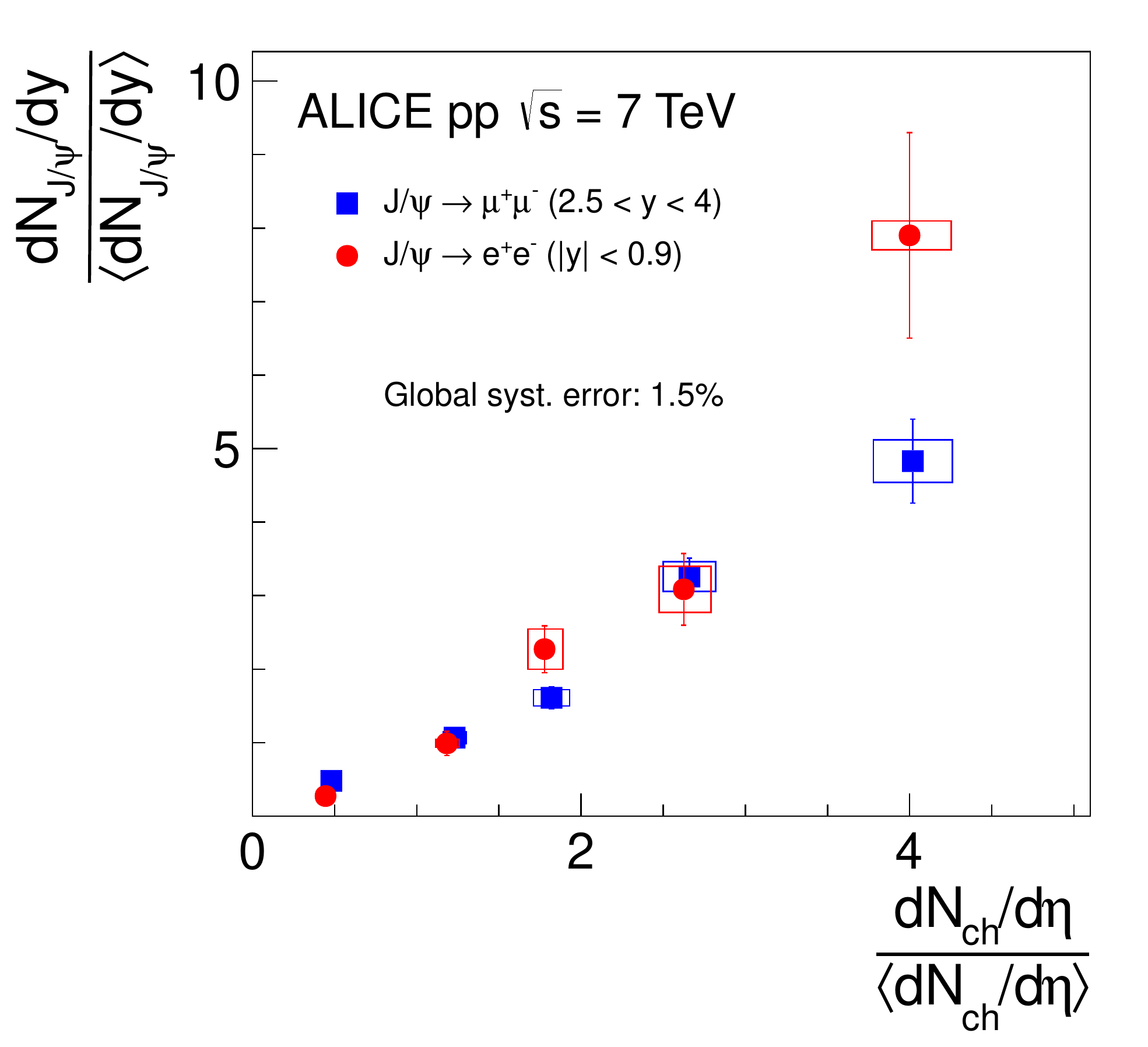}
\caption{\label{JPSIvsmult}d$N_{\textrm{J}/\psi}$/d$y$ as a function of the charged particle multiplicity density at mid-rapidity d$N_{\textrm{ch}}$/d$\eta$. Both values are normalized by the corresponding value for minimum bias pp collisions. Two measurements are shown: in red circles at mid-rapidity (J/$\psi \to e^+e^-$, $|y|<0.9$) and in blue squares at forward rapidities (J/$\psi \to \mu^+\mu^-$, $2.5<y<4$). Statistical uncertainties of the J/$\psi$ yields  are represented by error bars, while boxes reflect the quadratic sum of the point-to-point systematic uncertainties.} 
\end{center}
\end{figure}

For J/$\psi$ analysis in the di-muon channel, muon candidates are selected by requiring that at least one of the two muon candidates matches a trigger track reconstructed from at least three hits in the trigger chambers of the muon spectrometer. To remove muons produced at small angles that have crossed a significant fraction of the beam shield, a cut is applied on the radial coordinate of the track at the end of the front absorber ($R_{\textrm{abs}}>$17.5 cm). To reject events very close to the edge of the muon spectrometer acceptance a cut is applied on the rapidity of the pair ($2.5<y<4$). To obtain the number of J/$\psi$ in each multiplicty interval, a fit is used on the  corresponding di-muon invariant mass distribution in the range $2<M_{\textrm{inv}}<5$ GeV/$c^2$.  The line shapes of the J/$\psi$ and $\psi$(2S) are parametrised using a Crystal Ball function, while the underlying continuum is fitted with the sum of two exponential functions. Details on the quality of the fit results can be found in \cite{alice_Jpsi}. The two right panels of Fig. \ref{Minv} show the measured di-muon invariant mass distributions together with the results of the fit procedure for the lowest and highest multiplicity intervals.

\smallskip
Fig.~\ref{JPSIvsmult} presents the ratio of the J/$\psi$ yield in a given multiplicty interval relative to the minimum bias yield. Corrections regarding geometrical acceptance, reconstruction efficiency and their systematics cancel out in the ratio (d$N_{\textrm{J}/\psi}$/d$y$)/$<$d$N_{\textrm{J}/\psi}$/d$y>$. It was checked by Monte Carlo simulations that these corrections do not depend on d$N_{\textrm{ch}}$/d$\eta$, in the range under consideration (d$N_{\textrm{ch}}$/d$\eta < 32.9$). The number of events used for the normalization of $<$d$N_{\textrm{J}/\psi}/$d$y>$ is corrected for the fraction of inelastic events not seen by the trigger condition. After applying acceptance and efficiency corrections these values correspond to the values than can be extracted from data published in \cite{alice_Jpsi}:
$<$d$N_{\textrm{J}/\psi}/\textrm{d}y>=(8.2\pm0.8 \rm{ (stat.)}\pm1.2\rm{ (syst.)})\times10^5$ for J/$\psi \to e^+e^-$ in $|y| < 0.9$, and $<$d$N_{\textrm{J}/\psi}/\textrm{d}y>=(5.8\pm0.2 \rm{ (stat.)}\pm0.6\rm{ (syst.)})\times10^5$ for J/$\psi \to \mu^+ \mu^-$ in $2.5<y<4$.

\smallskip
For the di-electron analysis, the uncertainty due to background subtraction was obtained as the absolute differences using the   like-sign and the track rotation methods. It is found to be between 2\% and 12\% for the different multiplicity intervals. For the di-muon analysis this is evaluated by varying the functional form of the background description (polynomial instead of exponential). It depends on the signal to background ratio and varies between 3\% and 4\%. For the muon measurement an additional systematic uncertainty comes from pile-up events and is estimated to be 6\% in the first multiplicity interval and 3\% in the others. To account for the possible changes of the $p_{\rm{T}}$ spectrum with event multiplicity, an additional systematic uncertainty is determined by varying the $<p_{\rm{T}}>$ of the J/$\psi$ spectrum that is used as input to the determination of the Monte Carlo corrections between 2.6 and 3.2 GeV/c. A systematic uncertainty of 1.5\% (3.5\%) is found for the di-electron (di-muon) analysis. The total systematic error on (d$N_{\textrm{J}/\psi}$/d$y$)/$<$d$N_{\textrm{J}/\psi}$/d$y>$ is given by the quadratic sum of the different contributions and amounts to $2.5-12$ \% depending on the multiplicity interval for the di-electron result. In the case of the di-muon analysis, it varies between 8\% in the first and 6\% in the last multiplicity interval. An additional global uncertainty of 1.5\% on the normalization of  $<$d$N_{\textrm{J}/\psi}/$d$y>$ is introduced by the correction of the trigger inefficiency for all inelastic collisions. More detailed explanations on the estimation of systematic uncertainty estimated in this analysis can be found in \cite{jpsimult_paper}. 

In Fig.~\ref{JPSIvsmult} an approximately similar linear increase of the relative J/$\psi$ yield (d$N_{\textrm{J}/\psi}$/d$y$) / $<$dN$_{\textrm{J}/\psi}$/d$y>$ with (d$N_{\textrm{ch}}$/d$\eta$)/$<$d$N_{\textrm{ch}}$/d$\eta>$ is observed in both rapidity ranges. The enhancement relative to minimum bias J/$\psi$ yield is a factor of approximately 5 at $2.5 < y < 4$ (8 at $|y| < 0.9$) for 4 times the minimum bias charged particle multiplicity. A possible explanation for the observed correlation could be that J/$\psi$ is always accompanied by a strong hadronic activity biasing high multiplicity events. Such a mechanism could imply particular spatial distributions and J/$\psi$-hadron correlations could clarify the situation. Another possible mechanism would be initial density fluctuations accompanied by a specific structure of the nucleon. This mechanism seems to explain a factor 4 to 5 for charged multiplicity in J/$\psi$ events 5 times larger than in minimum bias events \cite{Strikman2}.

\section{First PYTHIA 6.4 comparison}

To compare our results to the predictions of a model, the model has to be able to reproduce all aspects of an event : the hard, the soft and semi-hard part of the event in a consistent framework. Event generators seem well suited to fulfill such a requirement. The considered event generator should also include heavy quarks ($c$ and $b$) and heavy resonances such as J/$\psi$ considering the correct masses of heavy quarks, energy conservation and color flow. Few models are left and none of them was built considering this new observable of J/$\psi$ yield versus multiplicity. The first one we can think of is PYTHIA 6.4 \cite{pythia6.4} which is a pp event generator commonly used at the LHC. This is not the best model adressing quarkonium production, but it is  extensively used, tuned and debugged. In this sense, this study is a first attempt of model-comparison for J/$\psi$ yield versus multiplicity.

\begin{figure}[!t]
\begin{center}
\includegraphics[scale=0.55]{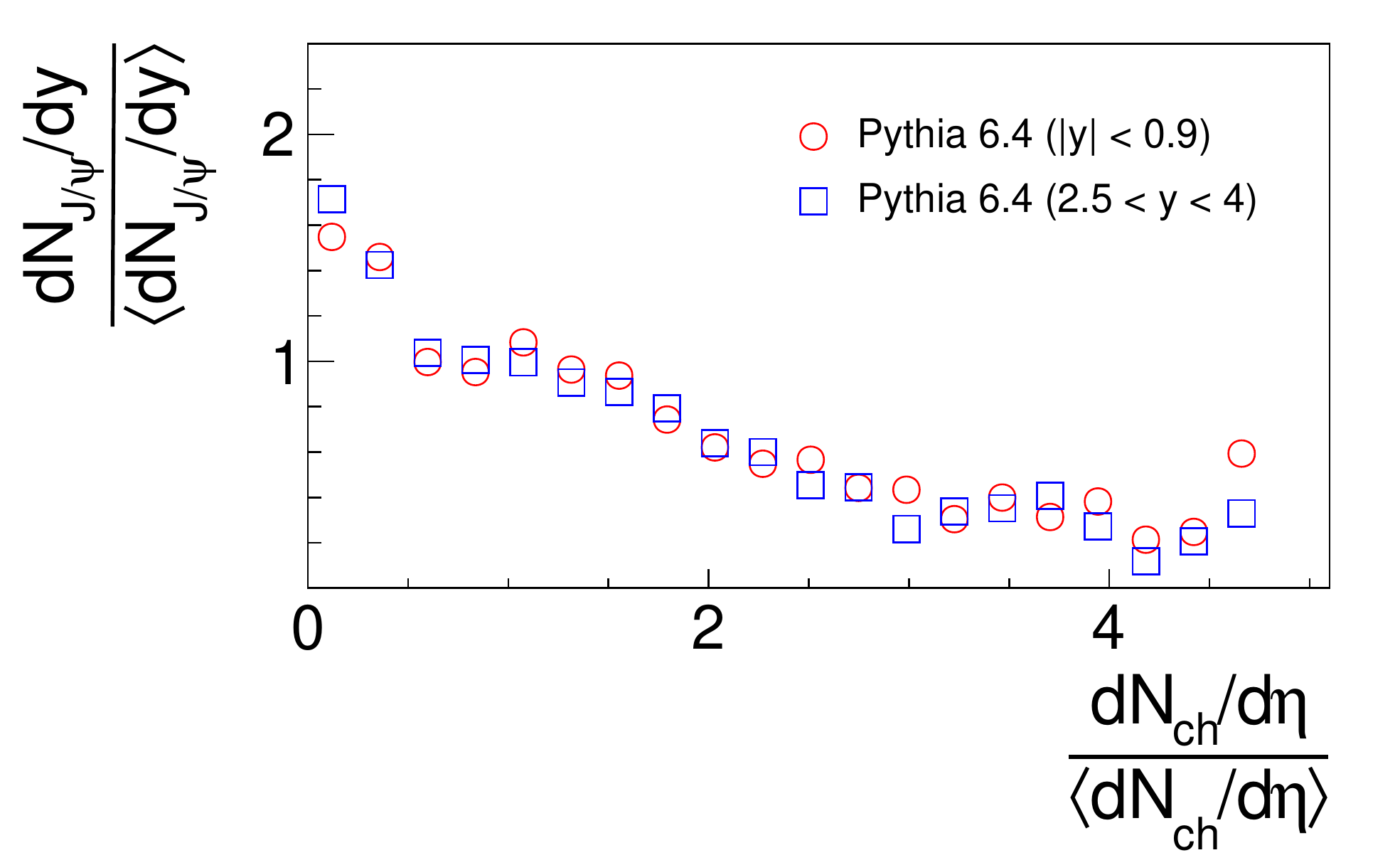}
\caption{\label{pythia-mstp350} Relative J/$\psi$ yield (d$N_{\textrm{J}/\psi}$/d$y$)/$<$d$N_{\textrm{J}/\psi}/$d$y>$ as a function of the relative charged particle multiplicity density around mid-rapidty (d$N_{\textrm{ch}}$/d$\eta$)/$<$d$N_{\textrm{ch}}$/d$\eta$ $>$ calculated with PYTHIA 6.4.25 in the PERUGIA 2011 tune MSTP(5)=350. Results are shown at forward ($2.5<y<4$) and central ($|y|<0.9$) rapidities for J/$\psi$ produced in hard scatterings via the NRQCD framework (MSEL=63).}
\end{center}
\end{figure}

\smallskip
Detailed explanations on the physics model of PYTHIA 6.4 can be found in \cite{pythia6.4}. A pp event is composed by a hard $2\to2$ process. The two partons from the incoming protons can evolve through initial state radiation processes (ISR) before the hard subprocess. The two produced partons can also evolve with final state radiation processes (FSR).  To produce J/$\psi$, several models are available, the color singlet (ISUB=86 with gluon fusion) where a J/$\psi$ is produced with a gluon to ensure momentum conservation, and NRQCD (ISUB=421-439) where a pre-resonant state is produced. This pre-resonant state will then decay into a J/$\psi$ emiting a soft gluon. In addition to the first hard interaction, a MPI scenario with varying impact parameter allows other incoming partons to undergo hard and semi-hard interactions. The first one is the hardest one. The following ones are ordered in hardness. In the last model of MPI implemented in PYTHIA 6.4, other interactions also evolve through ISR and FSR. In PYTHIA 6.4, quarkonia production in hard process is only available in the first hard interaction and not in the following ones of the MPI scenario. This was corrected in PYTHIA 8. In addition to the hard production, J/$\psi$ can be produced from B decays. This source can be easily turned off, imposing B hadrons to be stable. J/$\psi$ can also be produced by the fragmentation of a cluster made of a $c\bar{c}$ pair produced by the branching of one or two gluons into $c\bar{c}$ pairs. Those gluons come from ISR and FSR. At the end of the procedure all produced partons are connected with strings that will fragment into hadrons via the LUND string model \cite{pythia6.4}. 

\smallskip
In this study the version PYTHIA 6.4.25 is used with Perugia 2011 (P2011) tune, a recent tuning of PYTHIA including the fit of LHC pp data at 7 TeV. It is well suited to describe multiplicity distributions. We consider in this article only MSTP(5)=350 which is the major P2011 tune  \cite{Skands_perugia}. B hadrons are forced to be stable. 
Two different data samples are generated. The first one is minimum bias (MSEL=1) for the denominator of our observable. The second sample used in the denominator contains J/$\psi$. It is generated with MSEL=63 : J/$\psi$ are generated in hard processes with the NRQCD framework implemented in PYTHIA  \cite{pythia6.4}. This second sample also contains J/$\psi$ produced by cluster fragmentation. The identification of J/$\psi$ from hard processes and J/$\psi$ from cluster fragmentation can be done with the flavour code of the mother, allowing the separation of the two contributions. 
Finally, the same kinematical cuts  than in the ALICE experimental data sets are used to identify J/$\psi$ and charged particles.
As in the data analysis the samples are normalised by the number of events and the mean number of charged particles. The obtained results are shown in Fig.~\ref{pythia-mstp350} for PERUGIA 2011 tune with MSTP(5)=350. 
Results for J/$\psi$ produced in hard scattering only are presented for the same rapidity regions as in the data analysis.

\smallskip 
Concerning cluster contribution, further studies indicate that this contribution is strongly enhanced by the color reconnection mechanism. Clusters come from gluons originating from ISR and FSR. Therefore, this contribution should scale with multiplicity, because high multiplicity events correspond to events with a high partonic activity (MPI, ISR, FSR) and so an enhanced probability to form a cluster. 
Nevertheless this contribution in PYTHIA 6.4.25 does not seem to be under control due to the way color reconnection is implemented \cite{Skands_perugia} and warnings concerning the physics model of the cluster implementation can be found in the PYTHIA manual \cite{pythia6.4}. Thus, this contribution is excluded in results of Fig.~\ref{pythia-mstp350}.

\smallskip 
The J/$\psi$ production from hard processes only, decreases as a function of relative multiplicity, which is in disagreement with the trend observed in the data (Fig.~\ref{JPSIvsmult}). A na\"ive interpretation would be that if the first hard process is independent from the other aspects of the event (ISR, FSR, MPI, underlying event), the J/$\psi$ yields should be flat as a function of multiplicity. It is not observed with PYTHIA 6.4.25, and further studies with other models are needed to understand this behaviour. E.g. a study with PYTHIA 8 might provide further insight. In PYTHIA 8, MPI could play a role in J/$\psi$ production with the possibility of quarkonium production in all the parton-parton interactions of the MPI scenario. Other event generators, such as CASCADE \cite{cascade}, should be tested as well. 

\section{Conclusions}
In summary, we have presented the first measurement of the J/$\psi$ yield as a function of the charged particle multiplicity d$N_{\textrm{ch}}$/d$\eta$, performed by the ALICE collaboration \cite{jpsimult_paper}. 
J/$\psi$ mesons are detected at mid-rapidity ($|y| < 0.9$) and forward rapidity ($2.5 < y < 4$), while d$N_{\textrm{ch}}$/d$\eta$ is determined at mid-rapidity ($|\eta| < 1$). An approximately linear increase of the J/$\psi$ yields with the charged particle multiplicity is observed. The increase is similar at forward and mid-rapidity, exhibiting an enhancement relative to minimum bias J/$\psi$ yield by a factor of about 5 at $2.5 < y < 4$  (8 at $|y| < 0.9$) for $\sim 4$ times the minimum bias charged particle multiplicity. A first comparison with PYTHIA 6.4 simulations was shown. J/$\psi$ yields originated from the first hard interaction do not follow the same trend as seen in the data. Further studies are needed to explore the physics involved in such a new observable. From the experimental side the study of charged particle multiplicity dependence of $\Upsilon$, open charm and also other hard observables such as jet and Drell-Yan production could bring more informations. Multiplicity studies for various $p_{\rm{T}}$ classes are also of interest. One could also propose an underlying event study replacing the leading jet by a J/$\psi$ and J/$\psi$-hadrons correlation studies. From the event generators side, new studies are needed with PYTHIA 6.4, considering new tunes and other sets parameter.  Other models should also be tested such as PYTHIA 8 and CASCADE.


\begin{footnotesize}

\end{footnotesize}

\end{document}